\pdfoutput=1

\documentclass[12pt,a4paper]{article}

\usepackage{lineno}  
\usepackage{graphicx}  
\usepackage{xspace} 
\usepackage{color}
\usepackage{colortbl}
\usepackage{amsmath} 
\usepackage{ifthen} 

\newboolean{pdflatex}
\setboolean{pdflatex}{true} 
\newboolean{articletitles}
\setboolean{articletitles}{true} 

\newboolean{uprightparticles}
\setboolean{uprightparticles}{true} 

\usepackage{amssymb}
\usepackage{amsfonts}
\usepackage{upgreek} 

\usepackage{hyperref}    
\usepackage[all]{hypcap} 

\def\lhcb {\mbox{LHCb}\xspace}
\def\ux85 {\mbox{UX85}\xspace}

\def\atlas  {\mbox{ATLAS}\xspace}
\def\cms    {\mbox{CMS}\xspace}

\ifthenelse{\boolean{uprightparticles}}%
{

 \def\Pmu         {\ensuremath{\upmu}\xspace}                 
 \def\Pnu         {\ensuremath{\upnu}\xspace}                 
                  
 \def\Ppi         {\ensuremath{\uppi}\xspace}

 \def\Ptau        {\ensuremath{\uptau}\xspace}

 \def\PDelta      {\ensuremath{\Delta}\xspace}                 
 \def\PXi      {\ensuremath{\Xi}\xspace}                 
 \def\PLambda      {\ensuremath{\Lambda}\xspace}                 
 \def\PSigma      {\ensuremath{\Sigma}\xspace}                 
 \def\POmega      {\ensuremath{\Omega}\xspace}                 
 \def\PUpsilon      {\ensuremath{\Upsilon}\xspace}                 
 
 \def\PB      {\ensuremath{\mathrm{B}}\xspace}                 
                  
 \def\PD      {\ensuremath{\mathrm{D}}\xspace}

 \def\PK      {\ensuremath{\mathrm{K}}\xspace}

 \def\PW      {\ensuremath{\mathrm{W}}\xspace}

 \def\Pb      {\ensuremath{\mathrm{b}}\xspace}                 
 \def\Pc      {\ensuremath{\mathrm{c}}\xspace}                 
                  
 \def\Pe      {\ensuremath{\mathrm{e}}\xspace}

 \def\Pi      {\ensuremath{\mathrm{i}}\xspace}

 \def\Pp      {\ensuremath{\mathrm{p}}\xspace}

 \def\Pt      {\ensuremath{\mathrm{t}}\xspace}

}
{

 \def\Pmu         {\ensuremath{\mu}\xspace}                 
 \def\Pnu         {\ensuremath{\nu}\xspace}                 
                  
 \def\Ppi         {\ensuremath{\pi}\xspace}

 \def\Ptau        {\ensuremath{\tau}\xspace}

 \mathchardef\PDelta="7101
 \mathchardef\PXi="7104
 \mathchardef\PLambda="7103
 \mathchardef\PSigma="7106
 \mathchardef\POmega="710A
 \mathchardef\PUpsilon="7107
                  
 \def\PB      {\ensuremath{B}\xspace}                 
                  
 \def\PD      {\ensuremath{D}\xspace}

 \def\PK      {\ensuremath{K}\xspace}

 \def\PW      {\ensuremath{W}\xspace}

 \def\Pb      {\ensuremath{b}\xspace}                 
 \def\Pc      {\ensuremath{c}\xspace}                 
                  
 \def\Pe      {\ensuremath{e}\xspace}

 \def\Pi      {\ensuremath{i}\xspace}

 \def\Pp      {\ensuremath{p}\xspace}

 \def\Pt      {\ensuremath{t}\xspace}

}

\def\en         {\ensuremath{\Pe^-}\xspace}   
\def\ep         {\ensuremath{\Pe^+}\xspace}
\def\epm        {\ensuremath{\Pe^\pm}\xspace} 
\def\epem       {\ensuremath{\Pe^+\Pe^-}\xspace}

\def\mup        {\ensuremath{\Pmu^+}\xspace}
\def\mumu       {\ensuremath{\Pmu^+\Pmu^-}\xspace}

\def\tautau     {\ensuremath{\Ptau^+\Ptau^-}\xspace}

\def\neu        {\ensuremath{\Pnu}\xspace}

\def\neue       {\ensuremath{\neu_e}\xspace}

\def\neum       {\ensuremath{\neu_\mu}\xspace}

\def\Wp     {\ensuremath{\PW^+}\xspace}

\def\Wpm    {\ensuremath{\PW^\pm}\xspace}
\def\Z      {\ensuremath{\PZ^0}\xspace}

\def\cquark    {\ensuremath{\Pc}\xspace}

\def\bquark    {\ensuremath{\Pb}\xspace}

\def\tquark    {\ensuremath{\Pt}\xspace}
\def\tquarkbar {\ensuremath{\overline \tquark}\xspace}
\def\ttbar     {\ensuremath{\tquark\tquarkbar}\xspace}

\def\pion  {\ensuremath{\Ppi}\xspace}
\def\piz   {\ensuremath{\pion^0}\xspace}

\def\kaon  {\ensuremath{\PK}\xspace}
  \def\Kbar  {\kern 0.2em\overline{\kern -0.2em \PK}{}\xspace}

\def\Kz    {\ensuremath{\kaon^0}\xspace}
\def\Kzb   {\ensuremath{\Kbar^0}\xspace}
\def\KzKzb {\ensuremath{\Kz \kern -0.16em \Kzb}\xspace}
\def\Kp    {\ensuremath{\kaon^+}\xspace}
\def\Km    {\ensuremath{\kaon^-}\xspace}

\def\KpKm  {\ensuremath{\Kp \kern -0.16em \Km}\xspace}

  \def\Dbar    {\kern 0.2em\overline{\kern -0.2em \PD}{}\xspace}
\def\D       {\ensuremath{\PD}\xspace}

\def\Dz      {\ensuremath{\D^0}\xspace}
\def\Dzb     {\ensuremath{\Dbar^0}\xspace}
\def\DzDzb   {\ensuremath{\Dz {\kern -0.16em \Dzb}}\xspace}
\def\Dp      {\ensuremath{\D^+}\xspace}
\def\Dm      {\ensuremath{\D^-}\xspace}

\def\DpDm    {\ensuremath{\Dp {\kern -0.16em \Dm}}\xspace}

  \def\Bbar    {\kern 0.18em\overline{\kern -0.18em \PB}{}\xspace}

  \def\Y#1S{\ensuremath{\PUpsilon{(#1S)}}\xspace}

\def\proton      {\ensuremath{\Pp}\xspace}

\def\Lbar {\ensuremath{\kern 0.1em\overline{\kern -0.1em\PLambda}}\xspace}

\def\to                 {\ensuremath{\rightarrow}\xspace}

\def\order   {\ensuremath{\mathcal{O}}\xspace}

\newcommand{\as}{\ensuremath{\alpha_{\scriptscriptstyle S}}\xspace}

\def\AT#1     {\ensuremath{A_{\mathrm{T}}^{#1}}\xspace}           

\def\C#1      {\ensuremath{\mathcal{C}_{#1}}\xspace}                       
\def\Cp#1     {\ensuremath{\mathcal{C}_{#1}^{'}}\xspace}                    
\def\Ceff#1   {\ensuremath{\mathcal{C}_{#1}^{\mathrm{(eff)}}}\xspace}        
\def\Cpeff#1  {\ensuremath{\mathcal{C}_{#1}^{'\mathrm{(eff)}}}\xspace}       
\def\Ope#1    {\ensuremath{\mathcal{O}_{#1}}\xspace}                       
\def\Opep#1   {\ensuremath{\mathcal{O}_{#1}^{'}}\xspace}                    

\newcommand{\tev}{\ensuremath{\mathrm{\,Te\kern -0.1em V}}\xspace}
\newcommand{\gev}{\ensuremath{\mathrm{\,Ge\kern -0.1em V}}\xspace}
\newcommand{\mev}{\ensuremath{\mathrm{\,Me\kern -0.1em V}}\xspace}
\newcommand{\kev}{\ensuremath{\mathrm{\,ke\kern -0.1em V}}\xspace}
\newcommand{\ev}{\ensuremath{\mathrm{\,e\kern -0.1em V}}\xspace}
\newcommand{\gevc}{\ensuremath{{\mathrm{\,Ge\kern -0.1em V\!/}c}}\xspace}
\newcommand{\mevc}{\ensuremath{{\mathrm{\,Me\kern -0.1em V\!/}c}}\xspace}
\newcommand{\gevcc}{\ensuremath{{\mathrm{\,Ge\kern -0.1em V\!/}c^2}}\xspace}
\newcommand{\gevgevcccc}{\ensuremath{{\mathrm{\,Ge\kern -0.1em V^2\!/}c^4}}\xspace}
\newcommand{\mevcc}{\ensuremath{{\mathrm{\,Me\kern -0.1em V\!/}c^2}}\xspace}

\def\mum  {\ensuremath{\,\upmu\rm m}\xspace}

\def\pb {\ensuremath{\rm \,pb}\xspace}
\def\invpb {\ensuremath{\mbox{\,pb}^{-1}}\xspace}

\def\invfb   {\ensuremath{\mbox{\,fb}^{-1}}\xspace}

\def\order{{\ensuremath{\cal O}}\xspace}

\def\gsim{{~\raise.15em\hbox{$>$}\kern-.85em
          \lower.35em\hbox{$\sim$}~}\xspace}
\def\lsim{{~\raise.15em\hbox{$<$}\kern-.85em
          \lower.35em\hbox{$\sim$}~}\xspace}

\def\pt         {\mbox{$p_{\rm T}$}\xspace}
\def\et         {\mbox{$E_{\rm T}$}\xspace}

\newcommand{\lum} {\ensuremath{\mathcal{L}}\xspace}

\def\pythia     {\mbox{\textsc{Pythia}}\xspace}

\def\geant      {\mbox{\textsc{Geant4}}\xspace}

\def\tell1  {TELL1\xspace}
\def\ukl1   {UKL1\xspace}

\def\phistar {\mbox{$\phi^*$}}

\def\resbos {R{\sc esbos}}
\def\powheg {P{\sc owheg}}
\def\Z {\rm Z}

\usepackage{mciteplus}

\begin{document}


\begin{titlepage}
\pagenumbering{roman}

\vspace*{-1.5cm}
\centerline{\large EUROPEAN ORGANIZATION FOR NUCLEAR RESEARCH (CERN)}
\vspace*{1.5cm}
\hspace*{-0.5cm}
\begin{tabular*}{\linewidth}{lc@{\extracolsep{\fill}}r}
\ifthenelse{\boolean{pdflatex}}
{\vspace*{-2.7cm}\mbox{\!\!\!\includegraphics[width=.14\textwidth]{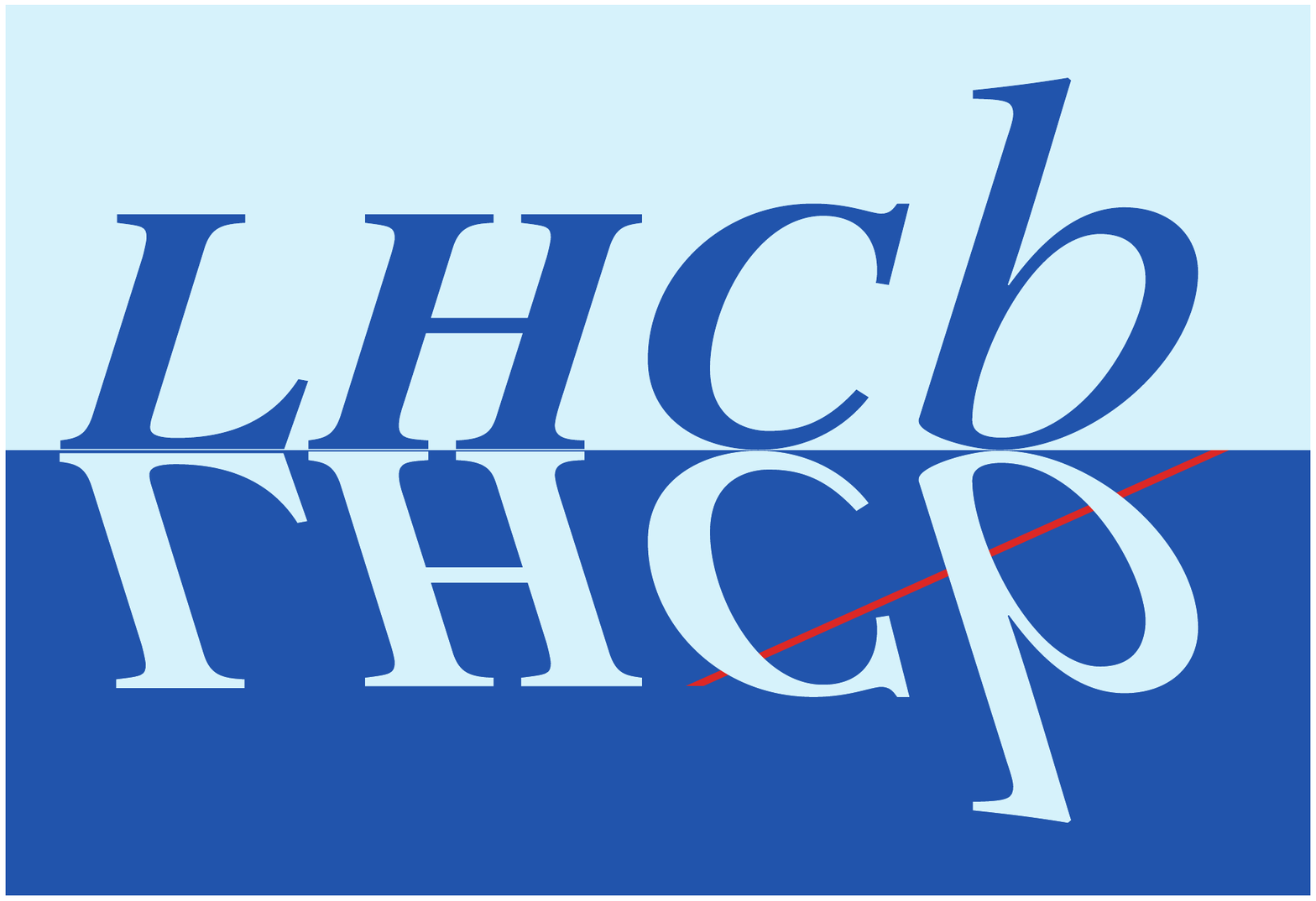}} & &}%
{\vspace*{-1.2cm}\mbox{\!\!\!\includegraphics[width=.12\textwidth]{lhcb-logo.eps}} & &}%
\\
 & & CERN-PH-EP-2012-363 \\  
 & & LHCb-PAPER-2012-036 \\  
 & & 21 January 2013
\end{tabular*}

\vspace*{4.0cm}

{\bf\boldmath\huge
\begin{center}
Measurement of the cross-section for $\Z\to\epem$ production in pp collisions at $\sqrt{s}=7$\tev

\end{center}
}

\vspace*{1.0cm}

\renewcommand{\thefootnote}{\fnsymbol{footnote}}
\addtocounter{footnote}{1}
\begin{center}
The LHCb collaboration\footnote{Authors are listed on the following pages.}
\end{center}
\renewcommand{\thefootnote}{\arabic{footnote}}
\addtocounter{footnote}{-2}
\
\vspace{\fill}

\begin{abstract}
  \noindent
A measurement of the cross-section for $\proton\proton\to\Z\to\epem$ is presented using
data at $\sqrt{s}=7$\tev corresponding to an integrated luminosity of 0.94\invfb.
The process is measured within the kinematic acceptance $\pt>20$\gevc and $2<\eta<4.5$ 
for the daughter electrons and dielectron invariant mass in the range \mbox{60--120\gevcc}. 
The cross-section is determined to be
$$\sigma(\proton\proton\to\Z\to\epem)=76.0\pm0.8\pm2.0\pm2.6\pb$$
where the first uncertainty is statistical, the second is systematic and the third is the uncertainty in the luminosity.
The measurement is performed as a function of \Z\ rapidity and as a function 
of an angular variable which is closely related to the \Z\ transverse momentum.
The results are compared with previous \lhcb\ measurements
and with theoretical predictions from QCD.
\end{abstract}

\vspace*{1mm}

\begin{center}
  Submitted to Journal of High Energy Physics
\end{center}

{\footnotesize 
\centerline{\copyright~CERN on behalf of the \lhcb collaboration, license \href{http://creativecommons.org/licenses/by/3.0/}{CC-BY-3.0}.}}

\vspace{\fill}

\end{titlepage}


\newpage
\setcounter{page}{2}
\mbox{~}
\newpage

\centerline{\large\bf LHCb collaboration}
\begin{flushleft}
\small
R.~Aaij$^{38}$, 
C.~Abellan~Beteta$^{33,n}$, 
A.~Adametz$^{11}$, 
B.~Adeva$^{34}$, 
M.~Adinolfi$^{43}$, 
C.~Adrover$^{6}$, 
A.~Affolder$^{49}$, 
Z.~Ajaltouni$^{5}$, 
J.~Albrecht$^{35}$, 
F.~Alessio$^{35}$
% ...

{\footnotesize \it
% ...
$ ^{5}$Clermont Universit\'{e}, Universit\'{e} Blaise Pascal, CNRS/IN2P3, LPC, Clermont-Ferrand, France\\
$ ^{6}$CPPM, Aix-Marseille Universit\'{e}, CNRS/IN2P3, Marseille, France\\
% ...
$ ^{11}$Physikalisches Institut, Ruprecht-Karls-Universit\"{a}t Heidelberg, Heidelberg, Germany\\
% ...
$ ^{33}$Universitat de Barcelona, Barcelona, Spain\\
$ ^{34}$Universidad de Santiago de Compostela, Santiago de Compostela, Spain\\
$ ^{35}$European Organization for Nuclear Research (CERN), Geneva, Switzerland\\
% ...
$ ^{38}$Nikhef National Institute for Subatomic Physics, Amsterdam, The Netherlands\\
% ...
$ ^{43}$H.H. Wills Physics Laboratory, University of Bristol, Bristol, United Kingdom\\
% ...
$ ^{49}$Oliver Lodge Laboratory, University of Liverpool, Liverpool, United Kingdom\\
% ...
$ ^{n}$LIFAELS, La Salle, Universitat Ramon Llull, Barcelona, Spain\\
% ...
}
\end{flushleft}

\cleardoublepage




\pagestyle{plain} 
\setcounter{page}{1}
\pagenumbering{arabic}


%

\section{Introduction}
\label{sec:Introduction}

The measurement of vector boson production permits a number of tests of electroweak physics and of quantum chromodynamics (QCD) to be performed.  In particular, the angular acceptance of \lhcb, roughly $2<\eta<5$ in the case of the main tracking system where $\eta$ denotes pseudorapidity, complements that of the general purpose detectors \atlas\ and \cms.
\lhcb\ measurements provide sensitivity to the proton structure functions at very low Bjorken $x$ values where the parton distribution functions (PDFs) are not particularly well constrained by previous data from HERA (see for example Ref.~\cite{Thorne:2008am}).

The most straightforward decay modes in which the \Wpm\ and \Z\ bosons can be studied using the \lhcb\ data 
are the muonic channels, 
{$\Z\to\mumu$} and {$\Wp\to\mup\neum$}.
Measurements of {$\Z\to\mumu$} and of {$\Z\to\tautau$} using the \lhcb\ data at $\sqrt{s}=7\tev$
have already been presented~\cite{LHCb-PAPER-2012-008,LHCb-PAPER-2012-029}.
To complement these studies, the electron channels
{$\Z\to\epem$} and {$\Wp\to\ep\neue$}, which offer statistically independent samples with
different sources of systematic uncertainties, are examined.

The main difficulty with electron\footnote{The term ``electron'' is used generically to refer to either \ep or \en.}  reconstruction in \lhcb\ is the energy measurement.  
A significant amount of material is traversed by the electrons before they reach the momentum analysing magnet, and their measured 
momenta are therefore liable to be reduced by bremsstrahlung.  For low energy electrons, the bremsstrahlung photons can frequently be identified in the electromagnetic calorimeter and their energies added to the measured momentum of the electron.  However, in the case of \Wpm\ and \Z\ decays, the electrons are of high momentum and transverse momentum (\pt), so that the bremsstrahlung photons often overlap with the electrons.  
The \lhcb\ calorimeters were designed so as to optimise the the measurement of
photons and $\piz$s from heavy flavour decays, whose transverse energy ($\et$) values are generally
well below 10 GeV.  As a consequence, individual calorimeter cells saturate at
\et\ around 10\gev, so it is not possible to substitute the calorimeter energy 
for the momentum measured using the spectrometer.
 We therefore have a situation in which the electron directions are well determined, but their energies are underestimated by a variable amount, typically around 25\%.  Nevertheless, the available information can be used to study certain interesting variables.

In this paper, we present a measurement of the cross-section for $\proton\proton\to\Z\to\epem$
using the data recorded by \lhcb\ in 2011 at $\sqrt{s}=7\tev$.
Throughout this paper we use $\Z\to\epem$ to refer to the process $\Z/\gamma^*\to\epem$ where either a virtual photon or a $\Z$ boson is produced and decays to $\epem$. For consistency, the measurement is presented in the same kinematic region as the recent measurement of {$\Z\to\mumu$} using the 2010 \lhcb\ data at $\sqrt{s}=7\tev$~\cite{LHCb-PAPER-2012-008}:
$2<\eta<4.5$ and $\pt>20\gevc$ for the leptons and $60<M<120$\gevcc for the dileptons where $M$ is the invariant mass.
Since the rapidity of the \Z\ boson can be determined 
to a precision of $\sim$0.05, the rapidity distribution will be presented.  
However, the \pt\ of the \Z\ boson is poorly determined and its distribution will not be discussed.
A similar problem was encountered by the D0 collaboration~\cite{Abazov:2010mk}, 
who employed a new variable proposed in Ref.~\cite{Banfi:2010cf} 
depending only on track angles
\begin{equation}
\phistar\equiv\tan\left(\frac{\phi_{\mathrm{acop}}}{2}\right)\left/\cosh\left(\frac{\Delta\eta}{2}\right)\approx\frac{\pt}{Mc}\,, \right.
\end{equation}
where $M$ and $\pt$ refer to the lepton pair, $\Delta\eta$ and $\Delta\phi$ are the differences in pseudorapidity and azimuthal angles respectively between the leptons, and the acoplanarity angle is $\phi_{\mathrm{acop}}=\pi-|\Delta\phi|$.
The \pt\ of the \Z\ boson is correlated with \phistar, and the resolution on \phistar\ is excellent, with a precision better than 0.001.  The measurement of \phistar\ presented here therefore largely accesses the same physics as a measurement of the \Z\ \pt\  distribution.
The measurement of the distribution of \Z\ rapidity (denoted $y_{\Z}$) is expected to show sensitivity to the choice of PDFs, while \phistar\ is likely to be more sensitive 
to higher order effects in the QCD modelling.

After a brief description of the detector, Sect.~\ref{sec:Selection} describes the event selection, and  Sect.~\ref{sec:Correction} outlines the determination of the cross-section.  The results are given in Sect.~\ref{sec:Results} followed by a short summary.

\section{\lhcb\ detector}
\label{sec:Detector}

The \lhcb detector~\cite{Alves:2008zz} is a single-arm forward
spectrometer covering the pseudorapidity range $2<\eta <5$, designed
primarily for the study of particles containing \bquark or \cquark quarks. The
detector includes a high precision tracking system consisting of a
silicon-strip vertex detector surrounding the pp interaction region,
a large-area silicon-strip detector located upstream of a dipole
magnet with a bending power of about $4{\rm\,Tm}$, and three stations
of silicon-strip detectors and straw drift tubes placed
downstream. The combined tracking system has a momentum resolution
$\Delta p/p$ that varies from 0.4\% at 5\gevc to 0.6\% at 100\gevc
for hadrons and muons,
and an impact parameter resolution of 20\mum for tracks with high
transverse momentum. Charged hadrons are identified using two
ring-imaging Cherenkov detectors. Photon, electron and hadron
candidates are identified by a calorimeter system consisting of
scintillating-pad (SPD) and preshower (PRS) detectors, an electromagnetic
calorimeter (ECAL) and a hadronic calorimeter (HCAL). The acceptance of the calorimeter 
system is roughly $1.8<\eta<4.3$. Muons are identified by a 
system composed of alternating layers of iron and multiwire
proportional chambers. 

The trigger~\cite{Aaij:2012me} consists of a hardware stage, based
on information from the calorimeter and muon systems, followed by a
software stage which applies full event reconstruction.
A significant improvement to the trigger was implemented during August 2011 which 
affected the trigger efficiency for \mbox{$\Z\to\epem$}.
The data samples before and after this change are treated separately and will be referred to as data sample I and data sample II.
These correspond to integrated luminosities of $581\pm20$\invpb\ and $364\pm13$\invpb\ respectively, yielding a total of $945\pm33$\invpb.

\section{Event selection}
\label{sec:Selection}

The $\Z\to\epem$ sample is initially selected by single-electron triggers, which 
require electrons
to have an $\et$ above a given threshold between 10 and 15\gev depending on the data-taking period and specific trigger.  
The $\Z\to\epem$ selection starts from a sample of \epem\ candidates with high invariant mass, 
which is refined by requiring the following selection criteria:
\begin{itemize}
\item At least one of the candidate electrons must be selected by a high-\et\ electron trigger.
\item The electrons are both required to have $\pt>20$\gevc and pseudorapidity in the range
$2.0<\eta<4.5$.
The invariant mass of the \epem\ pair should be greater than 40\gevcc.
\item Requirements on calorimeter information are imposed to provide particle identification (PID) of electrons. The particle must satisfy $E_{\mathrm{ECAL}}/pc>0.1$, where $p$ is the particle momentum, with bremsstrahlung correction if available, and $E_{\mathrm{ECAL}}$ is the ECAL energy associated with the particle.  The particle is required to lie within the HCAL acceptance and to satisfy $E_{\mathrm{HCAL}}/pc<0.05$, where $E_{\mathrm{HCAL}}$ is the HCAL energy associated with the particle.
The energy in the preshower detector associated with the particle is required to satisfy $E_{\mathrm{PRS}}>50$\mev.
These requirements impose an electromagnetic shower profile, while being loose enough to maintain a high electron efficiency despite the effects of calorimeter saturation and bremsstrahlung.
\item If more than one $\Z\to\epem$ candidate satisfies the above requirements in an event, just one candidate is used, chosen at random.
This only affects around 0.5\% of cases, and in all instances the multiple candidates share one daughter. 
\end{itemize}
A sample of same-sign $\epm\epm$ combinations, subject to the same selection criteria, is used to provide a data-based estimate of  background.  The main background is expected to arise from hadrons that 
shower early in the ECAL and consequently fake the signature of an electron.  These will contribute approximately equally to same-sign and opposite-sign pairs.  The contribution from semileptonic heavy flavour decays should be similar to the small level ($\sim0.2\%$) estimated for the $\Z\to\mumu$ channel~\cite{LHCb-PAPER-2012-008}; in any case, subtracting the same-sign contribution should account for most of this effect.  

Simulated event samples of $\Z\to\epem$ with $M(\epem)>40$\gevcc are also used to assess some efficiencies as discussed below. Simulated samples of $\Z\to\tautau$ and of \ttbar\ are used to assess possible background contributions.  
For the simulation, pp collisions are generated using
\pythia~6.4~\cite{Sjostrand:2006za} with a specific \lhcb
configuration~\cite{LHCb-PROC-2010-056} and the CTEQ6L1 PDF set~\cite{:2008zwNadolsky}.  
The interaction of the generated particles with the detector and its
response are implemented using the \geant
toolkit~\cite{Allison:2006ve, *Agostinelli:2002hh} as described in
Ref.~\cite{LHCb-PROC-2011-006}.
Simulated samples based on different versions of GEANT and of the detector model are
employed, which allows the reliability of the simulation to be assessed.
The simulated events are then reconstructed in the same way as the data, including simulation of the relevant trigger conditions.

\begin{figure}[tb]
  \begin{center}
    \includegraphics[scale=0.6]{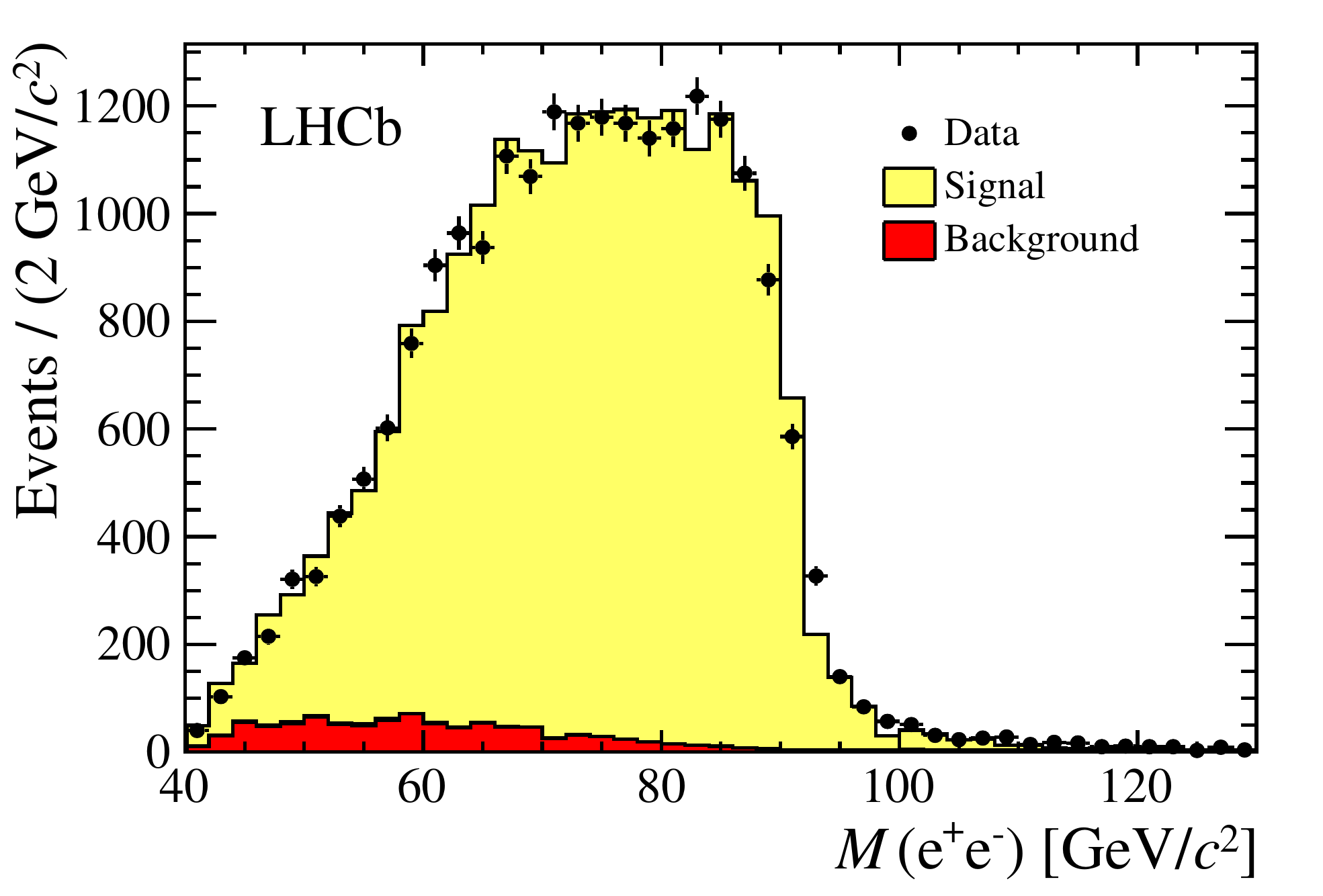}
    \vspace*{-1.0cm}
  \end{center}
  \caption{Invariant mass distribution of $\Z\to\epem$ candidates. The data are shown as points with error bars, the background obtained from same-sign data is shown in red (dark shading), to which the expectation from signal simulation is added in yellow (light shading).  
The $\Z\to\epem$ simulated distribution has been normalised to the (background-subtracted) data.}
  \label{fig:ZeeM} 
\end{figure}

The invariant mass distribution of the selected candidates is shown in Fig.~\ref{fig:ZeeM}.
The distribution falls off abruptly above the \Z\ mass
and is spread to lower masses by bremsstrahlung.  
Good agreement in shape is observed between data and the simulation sample used
in the data correction; this will be further discussed below. 
The background estimated from same-sign events amounts to 4.5\% of the total number of \epem\ candidates.  The backgrounds from 
\tautau\ and \ttbar\ events are estimated to be around 0.1\% and are neglected.

\section{Cross-section determination}
\label{sec:Correction}
In a given bin of \Z\ rapidity or \phistar, the cross-section is calculated using
\begin{equation}
 \sigma(\proton\proton\to\Z\to\epem) = \frac{N(\epem)-N(\epm\epm)}
 {\epsilon_{\mathrm{GEC}}\cdot
  \epsilon_{\mathrm{trig}}\cdot
  \epsilon_{\mathrm{track}}\cdot
  \epsilon_{\mathrm{kin}}\cdot
  \epsilon_{\mathrm{PID}}\cdot
   \int\lum\mathrm{d}t}
\cdot f_{\mathrm{FSR}}
\cdot f_{\mathrm{MZ}}
\;\;,
\label{equ:Corr}
\end{equation}
where $N(\epem)$ is the number of \Z\ candidates selected in data, $N(\epm\epm)$ is the background estimated from the number of same-sign candidates and $\int\lum\mathrm{d}t$ is the integrated luminosity.
The cross-section $\sigma(\proton\proton\to\Z\to\epem)$ denotes the product of the inclusive production cross-section 
for the $\Z$ or $\gamma^*$ and the branching ratio to $\epem$. 
The meaning and estimation of the other factors are described below.
The values obtained for each, averaged over the acceptance, are summarised in Table~\ref{tab:Effic}.

\begin{table}[t]
  \caption{Quantities entering into the cross-section determination, averaged over the range of \Z\ rapidity used.  
}
\begin{center}\begin{tabular}{l|cc}
                              & Data sample I & Data sample II \\ 
  \hline
$\int\lum\mathrm{d}t \; [\invpb]$ & $581\pm20$ & $364\pm13$ \\
$\epsilon_{\mathrm{GEC}}$ & \multicolumn{2}{c}{$0.947\pm0.004$} \\
$\epsilon_{\mathrm{trig}}$ & $0.715\pm0.021$ & $0.899\pm0.003$ \\
$\epsilon_{\mathrm{track}}$ & \multicolumn{2}{c}{$0.913\pm0.015$}\\
$\epsilon_{\mathrm{kin}}$ & \multicolumn{2}{c}{$0.500\pm0.007$}\\
$\epsilon_{\mathrm{PID}}$  & \multicolumn{2}{c}{$0.844\pm0.011$}\\
$f_{\mathrm{FSR}}$   & \multicolumn{2}{c}{$1.049\pm0.005$} \\
$f_{\mathrm{MZ}}$   &  \multicolumn{2}{c}{$0.967\pm0.001$}  \\ 
\end{tabular}\end{center}
\label{tab:Effic}
\end{table}

The luminosity is determined as described in Ref.~\cite{LHCb-PAPER-2011-015} 
and has an uncertainty of $3.5\%$.
The factor $f_{\mathrm{FSR}}$ accounts for the effects of final-state electromagnetic radiation, 
correcting the measurement to the Born level. As in the $\Z\to\mumu$ analysis~\cite{LHCb-PAPER-2012-008} it is determined
using P{\sc hotos}~\cite{Golonka:2005pn} interfaced to \pythia~\cite{Sjostrand:2006za}, 
with H{\sc orace}~\cite{CarloniCalame:2007cd} used as a cross-check.   An overall systematic uncertainty of 0.5\% is assigned to this correction~\cite{FSRsyst}.
The factor $f_{\mathrm{MZ}}$ corrects for $\epem$ events outside the mass range $60<M(\epem)<120$\gev which pass the event selection, and is estimated from simulation by examining the true mass for selected events.
 
The  probability for a $\Z\to\epem$ event to satisfy the trigger and selection requirements is given by the product of the efficiency factors, $\epsilon$, as described below.  
\begin{itemize}
\item 
Global event cuts (GEC) are applied in the trigger in order to prevent very large events from dominating the processing time.  Their efficiency for selecting signal events is given by $\epsilon_{\mathrm{GEC}}$.  In the $\Z\to\epem$ case, the most important requirement is on
the multiplicity of SPD hits, $N_{\mathrm{SPD}}\le 600$.  This is strongly correlated with the number of primary vertices reconstructed in the event.  The inefficiency is assessed by comparing 
with $\Z\to\mumu$ candidates recorded in the same running period using a dimuon trigger
for which a less stringent requirement of 900 hits is imposed.
A correction is made for the small difference in the numbers of
SPD hits associated with the electrons and muons themselves.
This procedure is adopted for each number of reconstructed primary vertices and the results are combined to obtain the overall efficiency.
\item The trigger efficiency for events passing the final selection, $\epsilon_{\mathrm{trig}}$, 
is determined from data.
A sample of events triggered independently of the \ep\ is identified  
and used to determine the efficiency for triggering the \ep, and likewise for the \en. 
Using the total numbers of candidates for which the single electron trigger is satisfied at each stage by the \ep\ ($N^+$), by the \en\ ($N^-$) and by both ($N^{+-}$), the efficiency for triggering the \ep\ is given by $\varepsilon^+=N^{+-}/N^-$.  
The overall efficiency is then taken to be $\varepsilon^-+\varepsilon^+-\varepsilon^-\varepsilon^+$ assuming 
that the \ep\ and \en\ are triggered independently.
The procedure is validated on simulated events.
The determination is performed separately in each bin of \Z\ rapidity and \phistar.
In all cases, the statistical uncertainty on the efficiency is taken as a contribution to the systematic uncertainty on the measurement.
\item
The track-finding efficiency, $\epsilon_{\mathrm{track}}$, represents the probability that both of the electrons are successfully 
reconstructed.  
The simulation is used to determine the track-finding efficiency, in bins of \Z\ rapidity and \phistar, by calculating the probability that, in a $\Z\to\epem$ event whose generated electrons lie within the kinematic acceptance, both of the electrons are associated with reconstructed tracks that satisfy the track quality requirements, but not necessarily the kinematic requirements.   Its statistical precision is propagated as a contribution to the systematic uncertainty.
\newline
This efficiency is checked in data using a tag-and-probe approach.  One electron is tagged 
using the standard requirements, and a search is made for an accompanying cluster of electromagnetic energy 
having a high \et\ and forming a high invariant mass with the tag electron.  If such a cluster has no associated track it provides evidence of a failure to reconstruct the other electron.  This sample contains significant background, which can be discriminated by examining the \pt\ distribution of the tag electron for cases where the photon candidate is and is not isolated.  The \pt\ distribution of the electrons in signal events in data displays a clear shoulder extending to $\sim$\,45\gevc 
while that for background falls monotonically, as shown in Fig.~\ref{fig:TrackEffFit}. The number of signal-like events in which a cluster is not associated with a track 
can be used to estimate a tracking efficiency, and the ratio of efficiencies between data and simulation is applied as a correction to the tracking efficiency.
The precision of the test is taken to define a systematic uncertainty, assumed to be fully correlated between bins of rapidity and \phistar.
\begin{figure}[tb]
  \begin{center}
    \includegraphics[scale=0.6]{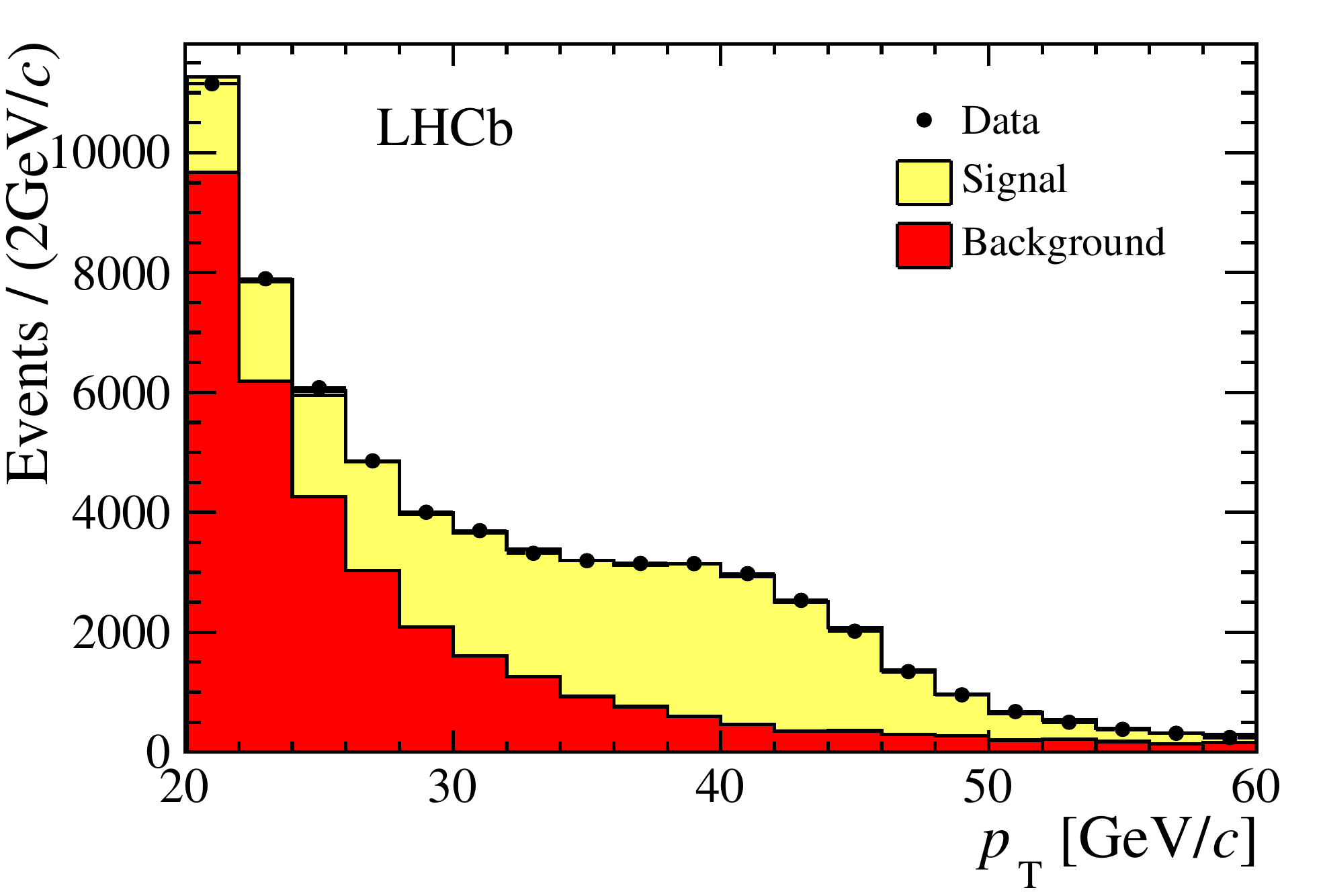}
    \vspace*{-1.0cm}
 \end{center}
  \caption{
Distribution of \pt\ for the ``tag'' electron in cases where an isolated cluster of energy of high \et\
is found in the electromagnetic calorimeter.  This is fitted with two components obtained from data, 
the $\Z\to\epem$ signal whose shape is taken from those candidates where the cluster is associated with an identified electron track, and background whose shape is obtained from candidates where the cluster is not isolated. 
}
  \label{fig:TrackEffFit}
\end{figure}
 
\item
The kinematic efficiency, $\epsilon_{\mathrm{kin}}$, represents the probability that, in a \mbox{$\Z\to\epem$} event whose generated electrons lie within the kinematic acceptance and are associated with reconstructed tracks, both tracks pass the kinematic selection requirements $2<\eta<4.5$ and \mbox{$\pt>20$\gevc}.
The efficiency is estimated from simulation, with its statistical precision being treated as a contribution to the systematic uncertainty.
\newline
This determination relies on a correct simulation, which can be tested using data.  For example, underestimation of the amount of material in the simulation would cause
a discrepancy between data and simulation 
in the \pt\ distributions of the electrons or the reconstructed mass spectrum shown in Fig.~\ref{fig:ZeeM}.
By comparing the shapes of the reconstructed mass spectrum and other kinematic distributions in data with different simulation samples, a systematic uncertainty on the momentum scale and hence on the kinematic efficiency is assigned.  This is combined with the statistical uncertainty mentioned above, with the systematic contribution taken to be fully correlated between bins of rapidity and \phistar.
\item
The PID efficiency, $\epsilon_{\mathrm{PID}}$, represents the probability that, in a $\Z\to\epem$ event with reconstructed electron tracks satisfying the kinematic requirements,
both tracks fulfil the calorimeter energy requirements for identified electrons.
This includes the probability that the tracks are within the calorimeter acceptance and have been successfully associated with calorimeter information.  Because of the acceptance contribution, the efficiency has a strong dependence on the \Z\ rapidity.  This dependence is taken from simulation, while the overall normalisation of the PID efficiency is estimated directly from data, using a tag-and-probe method. 
\newline
Starting from a sample which requires just one high \pt\ electron, events are selected by applying the usual
criteria except that only one of the \ep\ and \en\ (the ``tag'') is required to pass the calorimeter-based electron identification requirements.  
The other track is used as a ``probe'' to test the PID efficiency.  The requirement of only one identified electron admits a significant level of background, 
which is assessed similarly to the tracking efficiency by examining the \pt\ distribution of the tag or alternatively the \pt\ of the probe electron or the invariant mass of the two particles.  The size of the signal component can be used to 
define the number of \Z\ events which fail the PID, and hence to determine the PID efficiency and its uncertainty.
\end{itemize}

A systematic uncertainty is also assigned to the same-sign background subtraction. The assumption that same-sign $\epm\epm$ combinations model background in $\epem$ events is tested by selecting events which satisfy all criteria except that one of the particles fails the calorimeter energy requirements.  This sample should be dominated by background, and shows an excess of $\sim$8\% of opposite-sign events over same-sign events. Accordingly a systematic uncertainty amounting to 8\% of the number of same-sign events is assigned to the measurements.

\

\section{Results}
\label{sec:Results}

Using the efficiencies described above, the event yields detailed in Table~\ref{tab:Results} and Eq.~(\ref{equ:Corr}) separate cross-section measurements for the two data-taking periods are obtained.
Since these are in good agreement,
the results are combined using a weighted average, and assuming their uncertainties are fully correlated apart from 
the statistical contribution and the uncertainty in the trigger efficiency.  Data sample II has a smaller integrated luminosity but a higher and more precisely estimated trigger efficiency.  
The weighting of the two samples
is chosen to minimise the total uncertainty on the cross-section integrated over \Z\ rapidity.  
The values of the differential cross-sections obtained are given in Table~\ref{tab:Results}.  Correlation matrices may be found in the Appendix.
The bin $4.25<y_{\Z}<4.5$ is empty in data, and is expected to have close to zero detection efficiency since the calorimeter acceptance extends only slightly beyond 4.25.
Hence no measurement is possible.  However, the QCD calculations discussed below 
predict a cross-section below $\sim$0.01\pb in this bin, which is negligibly small, so comparisons with the 
\Z\to\mumu\ results or with theoretical calculations in the range $2<y_{\Z}<4.5$ are still meaningful.
 
The cross-section integrated over \Z\ rapidity is obtained by summing the cross-sections of all bins of $y_{\Z}$, taking the uncertainties associated with the GEC  and the luminosity to be fully correlated between bins, along with parts of the tracking, kinematic and PID efficiencies, and treating the other contributions as uncorrelated.
The cross-section is measured to be
 $$\sigma(\proton\proton\to\Z\to\epem)=76.0\pm0.8\,(\mathrm{stat.})\pm2.0\,(\mathrm{syst.})\pm2.6\,(\mathrm{lumi.})\pm0.4\,(\mathrm{FSR})\pb,$$ 
where the first uncertainty is statistical, the second is the experimental systematic uncertainty, 
the third is the luminosity uncertainty and the last represents the uncertainty in the FSR correction.
Since the results have been corrected to the Born level using the factor $f_{\mathrm{FSR}}$, it is possible to compare this measurement with that found in the \Z\to\mumu\ analysis~\cite{LHCb-PAPER-2012-008} using 37\invpb of data, namely $76.7\pm1.7\,(\mathrm{stat.})\pm3.3\,(\mathrm{syst.})\pm2.7\,(\mathrm{lumi.})$~pb.
Accounting for correlated uncertainties, the ratio of cross-sections is
$$ \frac{\sigma(\proton\proton\to\Z\to\epem)}{\sigma(\proton\proton\to\Z\to\mumu)}=0.990\pm0.024\,(\mathrm{stat.})\pm0.044\,(\mathrm{syst.}).$$
This may be regarded as a cross-check of the analyses.
Assuming lepton universality, the two cross-sections can be combined in a weighted average so as to minimise the total uncertainty, yielding
 $$\sigma(\proton\proton\to\Z\to\ell^+\ell^-)=76.1\pm0.7\,(\mathrm{stat.})\pm1.8\,(\mathrm{syst.})\pm2.7\,(\mathrm{lumi.})\pm0.4\,(\mathrm{FSR})\pb.$$ A recent measurement in $\Z\to\tautau$ decays which has a larger statistical uncertainty~\cite{LHCb-PAPER-2012-029}  can
also be combined with the electron and muon channels, yielding
 $$\sigma(\proton\proton\to\Z\to\ell^+\ell^-)=75.4\pm0.8\,(\mathrm{stat.})\pm1.7\,(\mathrm{syst.})\pm2.6\,(\mathrm{lumi.})\pm0.4\,(\mathrm{FSR})\pb.$$ 

\begin{table}[t]
  \caption{Event yields and measurements for the differential cross-section of
\mbox{$\proton\proton\to\Z\to\epem$} at $\sqrt{s}=$7~TeV as a function of \Z\ rapidity, $y_{\Z}$, and of \phistar.
The first uncertainty is statistical, the second and third are the uncorrelated and correlated experimental systematic uncertainties respectively, and the fourth is the uncertainty in the FSR correction. The common luminosity uncertainty of 3.5\% is not explicitly included here. 
The results are given for the combined data sample.
The right-hand column gives the values used for the FSR correction factor. 
}
\begin{center}
\scalebox{1.}{
\begin{tabular}{r@{--}l|r|r|c|c}
\multicolumn{2}{c|}{$y_{\Z}$}  & $N(\epem)$ & $N(\epm\epm)$&  d$\sigma$/d$y_{\Z}$ [pb] & $f_{\mathrm{FSR}}$ \\ \hline
2.00 & 2.25 & 988 & 40     & $13.6\pm0.7\pm0.4\pm0.3\pm0.1$ & $1.049\pm0.004$ \\
2.25 & 2.50 & 3064 & 121 & $39.4\pm1.0\pm0.6\pm0.8\pm0.2$  & $1.046\pm0.002$\\
2.50 & 2.75 & 4582 & 202 & $56.7\pm1.2\pm0.7\pm1.3\pm0.3$ & $1.050\pm0.002$ \\
2.75 & 3.00 & 5076 & 214   & $63.2\pm1.3\pm0.8\pm1.5\pm0.3$ & $1.049\pm0.002$ \\
3.00 & 3.25 & 4223 & 181   & $59.9\pm1.4\pm0.8\pm1.6\pm0.3$ & $1.056\pm0.002$ \\
3.25 & 3.50 & 2429 & 135 & $43.8\pm1.3\pm0.8\pm1.1\pm0.2$  & $1.054\pm0.003$\\
3.50 & 3.75 & 906 & 61   & $20.5\pm1.0\pm0.7\pm0.6\pm0.1$  & $1.030\pm0.006$\\
3.75 & 4.00 & 143 & 18     & $ 5.9\pm0.8\pm0.5\pm0.3\pm0.1$  & $1.074\pm0.029$\\
4.00 & 4.25 & 9 & 2       & $0.66\pm0.44\pm0.30\pm0.04\pm0.02$  & $1.074\pm0.029$\\
4.25 & 4.50 & 0 & 0 &  --- & \\
\end{tabular}
}
\vspace{10mm}
\scalebox{1.}{
\begin{tabular}{r@{--}l|r|r|c|c}
\multicolumn{6}{c}{}  \\
\multicolumn{2}{c|}{$\phistar$}  & $N(\epem)$ & $N(\epm\epm)$&  d$\sigma$/d$\phistar$ [pb]  & $f_{\mathrm{FSR}}$\\ \hline
0.00 & 0.05 & 9696 & 363  & $693\pm10\pm6\pm17\pm3$  & $1.059\pm0.001$\\
0.05 & 0.10 & 4787 & 219 & $326\pm7\pm4\pm8\pm2$  & $1.047\pm0.002$\\
0.10 & 0.15 & 2382 & 115 & $164\pm5\pm3\pm4\pm1$  & $1.039\pm0.002$\\
0.15 & 0.20 & 1384 & 80 & $99.1\pm4.0\pm2.0\pm2.2\pm0.5$  & $1.043\pm0.003$\\
0.20 & 0.30 & 1434 & 82 & $49.6\pm2.0\pm1.1\pm1.0\pm0.3$  & $1.042\pm0.003$\\
0.30 & 0.40 & 707 & 39  & $25.5\pm1.4\pm0.8\pm0.6\pm0.1$  & $1.049\pm0.004$\\
0.40 & 0.60 & 583 & 41  & $10.8\pm0.7\pm0.4\pm0.3\pm0.1$  & $1.052\pm0.005$\\
0.60 & 0.80 & 217 & 13  & $4.05\pm0.38\pm0.20\pm0.09\pm0.03$  & $1.054\pm0.005$\\
0.80 & 1.00 & 91 & 9 &     $1.41\pm0.23\pm0.11\pm0.03\pm0.02$  & $1.051\pm0.009$\\
1.00 & 2.00 & 119 & 9 &     $0.41\pm0.06\pm0.03\pm0.01\pm0.02$  & $1.035\pm0.011$\\
\end{tabular}
}
\end{center}
\label{tab:Results}
\end{table}

\begin{figure}[tb]
  \begin{center}
\includegraphics[scale=0.6]{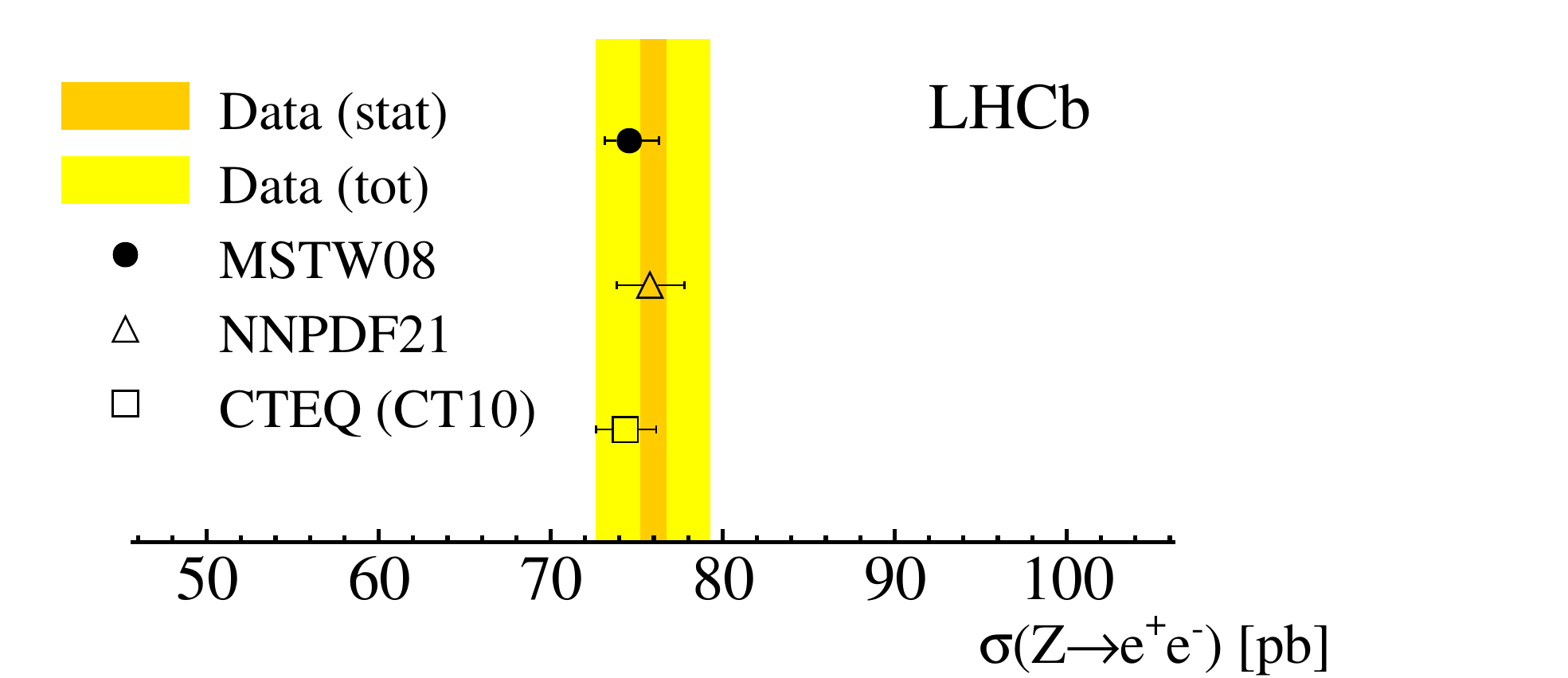}
    \vspace*{-0.5cm}
 \end{center}
  \caption{Cross-section for $\proton\proton\to\Z\to\epem$ at $\sqrt{s}=7\tev$ measured in \lhcb, shown as the yellow band.
The inner (darker) band represents the statistical uncertainty and the outer the total uncertainty.
 The measurement corresponds to the kinematic acceptance,  $\pt>20$\gevc and $2<\eta<4.5$ for the leptons  and $60<M<120$\gevcc for the dilepton.  
The points show the various theoretical predictions with their uncertainties as described in the text.}
  \label{fig:XSplot}
\end{figure}

\begin{figure}[tbh]
  \begin{center}
    \vspace*{-1.0cm}
    \includegraphics[scale=0.39]{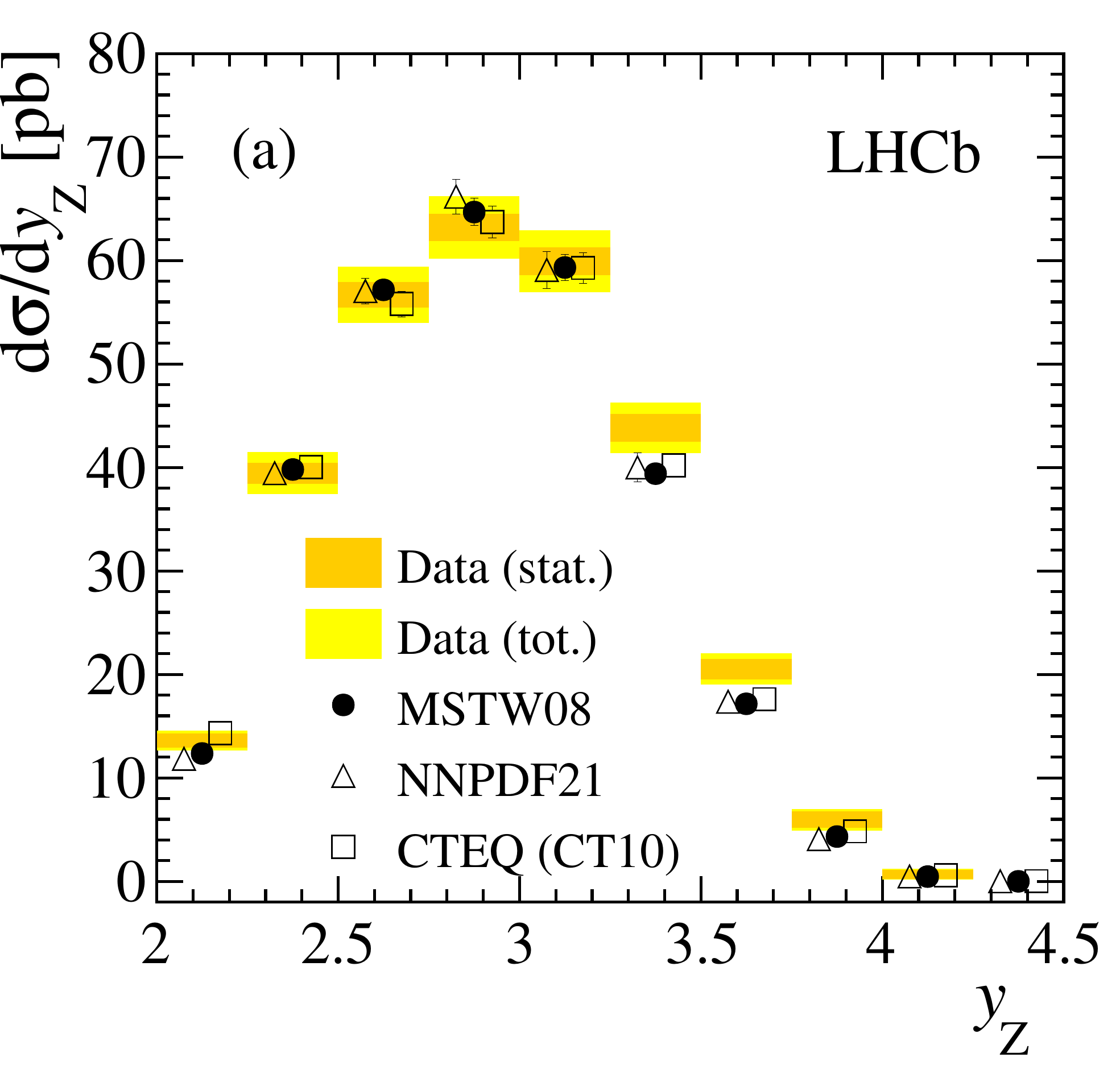}
    \includegraphics[scale=0.39]{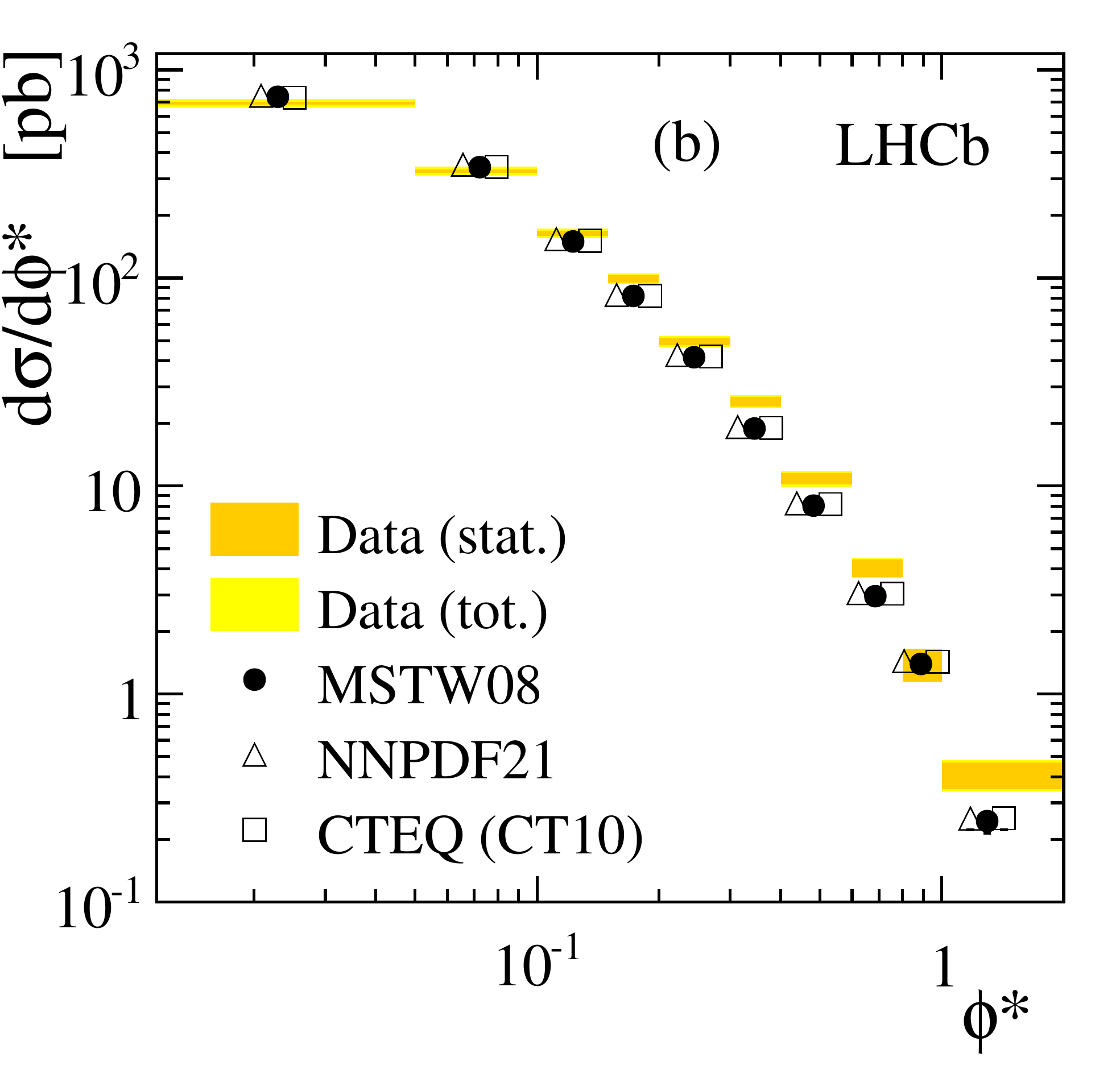}
    \vspace*{-1.0cm}
 \end{center}
  \caption{
Differential cross-section for $\proton\proton\to\Z\to\epem$ as a function of (a) \Z\ rapidity and  
(b) \phistar.
The measurements based on the $\sqrt{s}=7\tev$ \lhcb\ data are shown as the yellow bands where
the inner (darker) band represents the statistical uncertainty and the outer the total uncertainty.  NNLO QCD predictions are shown as points with error bars reflecting their uncertainties as described in the text.
}
  \label{fig:zeta}
\end{figure}

The results may be compared with theoretical calculations similar 
to those used in the interpretation of the $\Z\to\mumu$ analysis~\cite{LHCb-PAPER-2012-008}.
These calculations are performed at NNLO ($\order(\as^2)$) with the program 
FEWZ~\cite{Gavin:2010az} version 2.1.1 and using
the NNLO PDF sets of 
MSTW08~\cite{Martin:2009iq}, 
NNPDF21~\cite{Ball:2010de} or CTEQ (CT10~NNLO)~\cite{Lai:2010vv,*Nadolsky:2012ia}.
In Fig.~\ref{fig:XSplot} we present the measured cross-section and 
in Fig.~\ref{fig:zeta}(a) the measurements of the \Z\ rapidity distribution, compared in each case with the three calculations.  The uncertainties in the predictions include the effect of varying the renormalisation and factorisation scales by factors of two around the nominal value, which is set to the \Z\ mass, 
combined in quadrature with the PDF uncertainties at 68\% confidence level. 
The data agree with expectations within the uncertainties.

\begin{figure}[tb]
  \begin{center}
    \vspace*{-1.0cm}
    \includegraphics[scale=0.39]{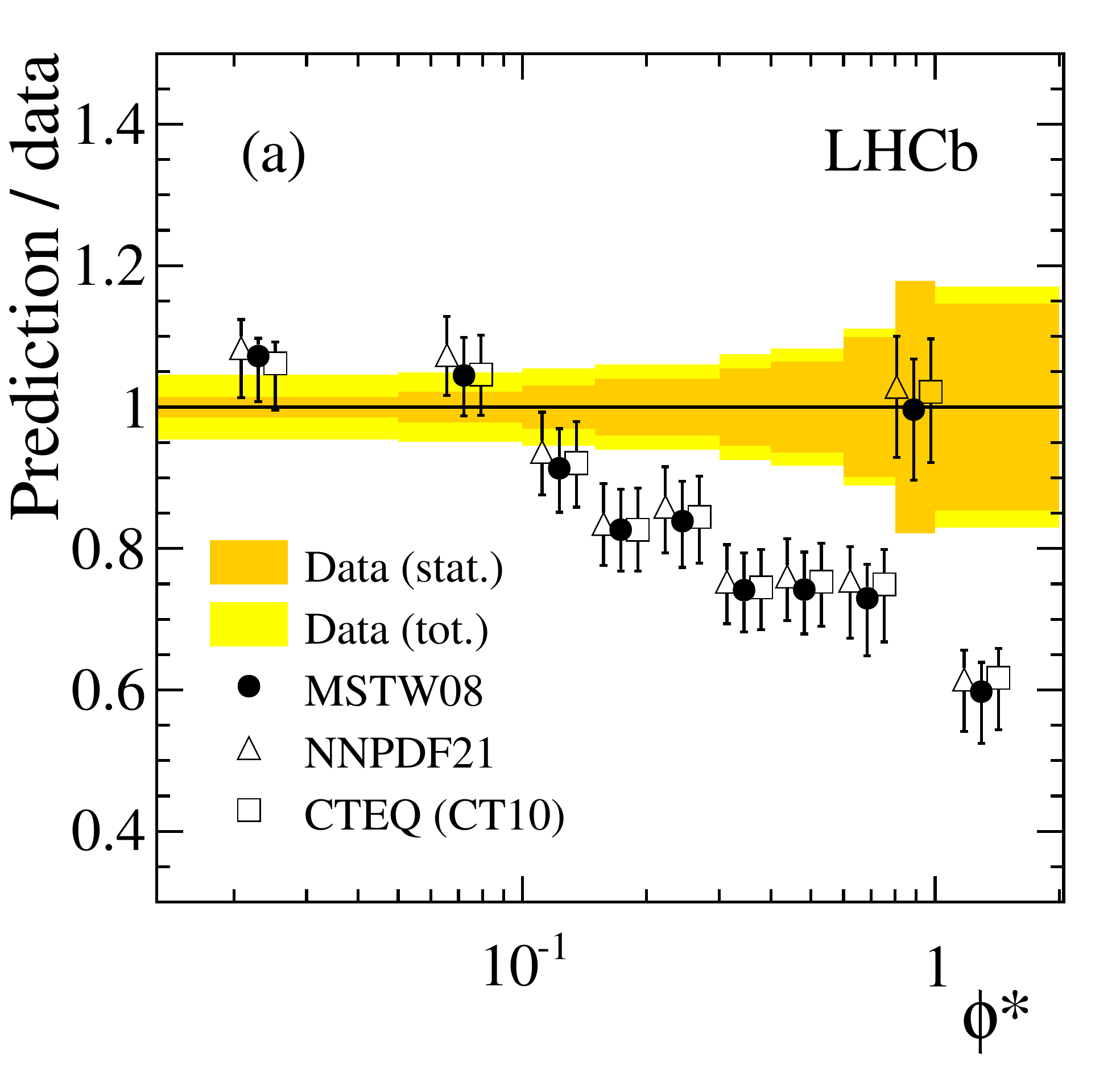}
    \includegraphics[scale=0.39]{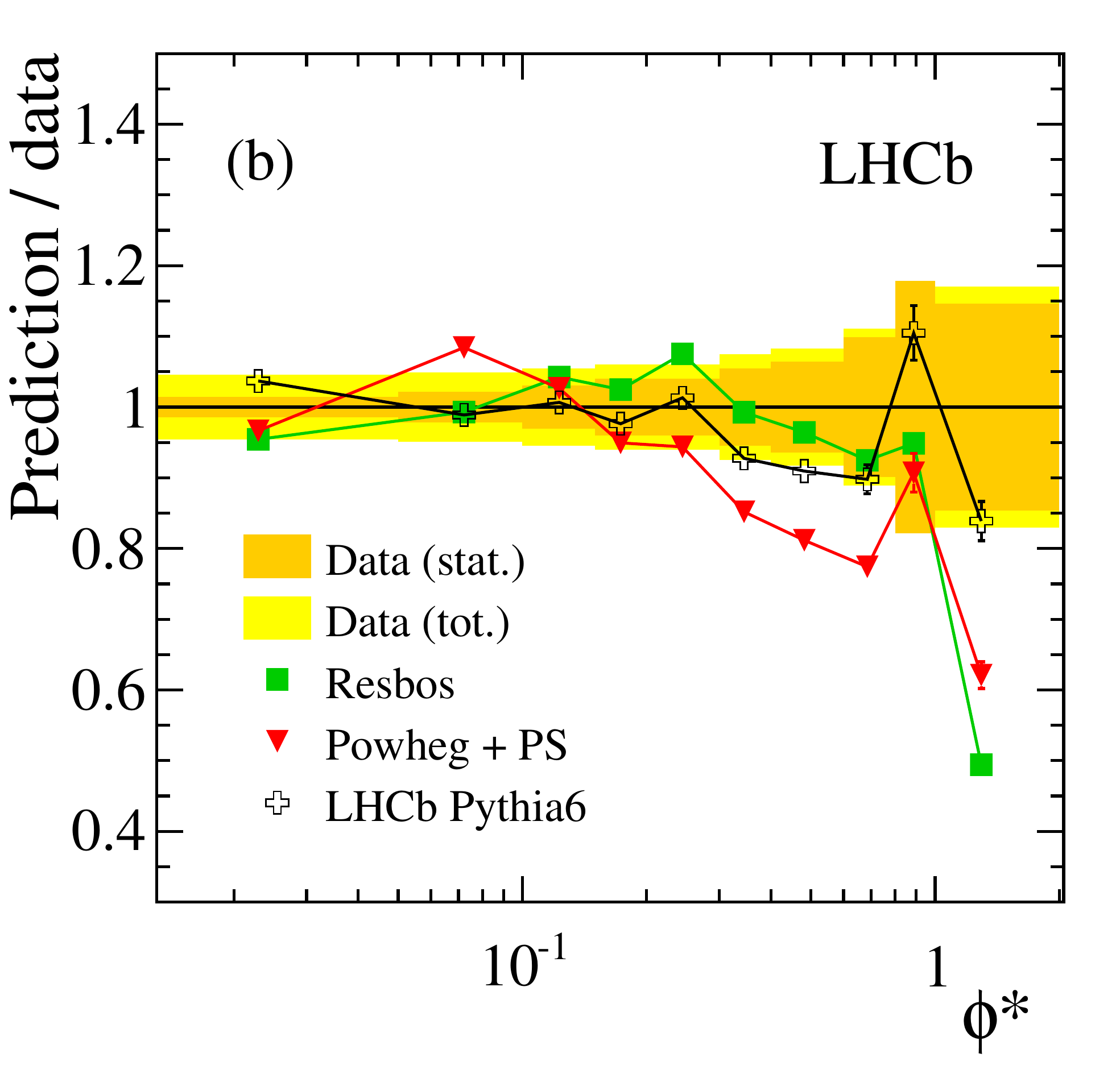}
    \vspace*{-1.0cm}
 \end{center}
  \caption{
Ratios of various QCD calculations to data for the differential cross-section for $\proton\proton\to\Z\to\epem$ as a function of \phistar.  The measurements based on the $\sqrt{s}=7\tev$ \lhcb\ data are shown as the yellow band centred at unity where the inner (darker) band represents the statistical uncertainty and the outer the total uncertainty.
(a) NNLO QCD predictions shown as points with error bars reflecting 
their uncertainties as described in the text.  Small lateral displacements of the theory points are made to improve clarity.
(b) Ratios of the predictions of \pythia, 
\resbos\ and \powheg\ to the data shown as points, with error bars that reflect the statistical uncertainties in the predictions.  For most points, these errors are so small that they are not visible.
}
\label{fig:zphistar2}
\end{figure}

The differential cross-section as a function of \phistar\ is shown in Fig.~\ref{fig:zeta}(b),  
compared with the predictions of QCD to NNLO.  Figure~\ref{fig:zphistar2}(a) displays 
the ratios of these predictions to the measurements. 
The NNLO calculations tend to overestimate the data at low \phistar\ and to underestimate the data at high \phistar.  It is expected that the \phistar\ distribution, like that of \pt, is significantly affected by multiple soft gluon emissions, which are not sufficiently accounted for in fixed order calculations.
A QCD calculation which takes this into account through resummation is provided by \resbos~\cite{Ladinsky:1993zn,*Balazs:1997xd,*Landry:2002ix}.\footnote{The P branch of \resbos\ is used with 
grids for LHC at $\sqrt{s}=7\tev$ based on CTEQ6.6.}
Another resummed calculation~\cite{Banfi:2012du} has been compared with ATLAS data~\cite{AtlasPhistar} in the central region of rapidity, but is not yet available for the \lhcb\ acceptance. 
Alternatively, \powheg~\cite{Alioli:2008gx,*Alioli:2010qp} 
provides a framework whereby a NLO QCD ($\order(\as)$)
calculation can be interfaced to a 
parton shower model such as \pythia\ which can approximate higher order effects.
Comparisons with these models, and with the \lhcb\ version~\cite{LHCb-PROC-2010-056} of \pythia~\cite{Sjostrand:2006za} are shown in Fig.~\ref{fig:zphistar2}(b).  The \resbos\ and \powheg\ distributions are normalised to their own cross-section predictions, while the \pythia\ distribution is normalised to the cross-section measured in data. It is seen that \resbos\ gives a reasonable description of the \phistar\ distribution. 
\powheg\ shows that the combination of a parton shower with the $\order(\as)$ QCD prediction  significantly improves the description of data in the low \phistar\ region, while in the high \phistar\ region the data are still underestimated.  
\pythia\ models the data reasonably well.  Overall, \resbos\ and \pythia\ seem to be the more successful of the calculation schemes considered here. 

\clearpage

\section{Summary}
\label{sec:Summary}

A measurement of the $\proton\proton\to\Z\to\epem$ cross-section in 
pp collisions at $\sqrt{s}=7$\tev using 0.94\invfb\ of data recorded by \lhcb\ is presented. 
Although the characteristics of the \lhcb\ detector prevent
a sharp mass peak from being seen, a clean sample of 
events is identified with less than 5\% background. 
Within the kinematic acceptance, $\pt>20$\gevc and $2<\eta<4.5$ for the leptons and $60<M<120$\gevcc
for the dielectron, the cross-section is measured to be 
 $$\sigma(\proton\proton\to\Z\to\epem)=76.0\pm0.8\,(\mathrm{stat.})\pm2.0\,(\mathrm{syst.})\pm2.6\,(\mathrm{lumi.})\pm0.4\,(\mathrm{FSR})\pb.$$ 
The cross-section is also measured in bins of the rapidity of the \Z\ and of the angular variable \phistar. The measurements of the rapidity distribution and of the integrated cross-sections are consistent with previous measurements using \Z\ decays to \mumu\ and \tautau\ 
and show good agreement with the expectations from NNLO QCD calculations.  The \phistar\ distribution, related to the \Z\ \pt\ distribution, is better modelled by calculations which approximately include the effects of higher orders.

\section*{Acknowledgements}

\noindent We express our gratitude to our colleagues in the CERN
accelerator departments for the excellent performance of the LHC. We
thank the technical and administrative staff at the LHCb
institutes. We acknowledge support from CERN and from the national
agencies: CAPES, CNPq, FAPERJ and FINEP (Brazil); NSFC (China);
CNRS/IN2P3 and Region Auvergne (France); BMBF, DFG, HGF and MPG
(Germany); SFI (Ireland); INFN (Italy); FOM and NWO (The Netherlands);
SCSR (Poland); ANCS/IFA (Romania); MinES, Rosatom, RFBR and NRC
``Kurchatov Institute'' (Russia); MinECo, XuntaGal and GENCAT (Spain);
SNSF and SER (Switzerland); NAS Ukraine (Ukraine); STFC (United
Kingdom); NSF (USA). We also acknowledge the support received from the
ERC under FP7. The Tier1 computing centres are supported by IN2P3
(France), KIT and BMBF (Germany), INFN (Italy), NWO and SURF (The
Netherlands), PIC (Spain), GridPP (United Kingdom). We are thankful
for the computing resources put at our disposal by Yandex LLC
(Russia), as well as to the communities behind the multiple open
source software packages that we depend on.

\clearpage

\clearpage
\setcounter{table}{0}
\setcounter{section}{1}
\renewcommand\thetable{\Alph{section}.\arabic{table}}

\section*{Appendix}
\label{sect:Appendix}
\setcounter{table}{0}
\setcounter{section}{1}
\renewcommand\thetable{\Alph{section}.\arabic{table}}
\begin{table}[h]
  \caption{Correlation coefficients for the differential cross-section of $\Z\to\epem$ at 7~TeV between bins of \Z\ rapidity, $y_{\Z}$.  Both statistical and systematic contributions are included.}
\begin{center}
\scalebox{0.84}{
\begin{tabular}{r@{--}l|ccccccccc}
\multicolumn{2}{c|}{$y_{\Z}$ bin} 
& 2.--2.25 & 2.25--2.5 & 2.5--2.75 & 2.75--3. & 3.--3.25  & 3.25--3.5 & 3.5--3.75 & 3.75--4. & 4.--4.25 \\ \hline
2.00 & 2.25 & 1 \\
2.25 & 2.50 & 0.47 & 1 \\
2.50 & 2.75 & 0.50 & 0.70 & 1 \\
2.75 & 3.00 & 0.51 & 0.70 & 0.75 & 1 \\
3.00 & 3.25 & 0.50 & 0.69 & 0.74 & 0.75 & 1 \\
3.25 & 3.50 & 0.45 & 0.62 & 0.66 & 0.67 & 0.66 & 1 \\
3.50 & 3.75 & 0.35 & 0.49 & 0.52 & 0.52 & 0.51 & 0.46 & 1 \\
3.75 & 4.00 & 0.20 & 0.27 & 0.29 & 0.29 & 0.29 & 0.26 & 0.20 & 1 \\
4.00 & 4.25 & 0.05 & 0.07 & 0.08 & 0.08 & 0.08 & 0.07 & 0.06 & 0.03 & 1 \\
\end{tabular}
}
\end{center}
\label{tab:CorrelY}
\end{table}

\begin{table}[h]
  \caption{Correlation coefficients for the differential cross-section of $\Z\to\epem$ at 7~TeV between bins of \phistar.  Both statistical and systematic contributions are included.}
\begin{center}
\scalebox{0.84}{
\begin{tabular}{r@{--}l|cccccccccc}
\multicolumn{2}{c|}{$\phistar$ bin} 
& 0.--0.05 & 0.05--0.1 & 0.1--0.15 & 0.15--0.2 & 0.2--0.3  & 0.3--0.4 & 0.4--0.6 & 0.6--0.8 & 0.8--1. & 1.--2. \\ \hline
0.00 & 0.05 & 1 \\
0.05 & 0.10 & 0.80 & 1 \\
0.10 & 0.15 & 0.73 & 0.67 & 1 \\
0.15 & 0.20 & 0.63 & 0.58 & 0.53 & 1 \\
0.20 & 0.30 & 0.62 & 0.58 & 0.53 & 0.45 & 1 \\
0.30 & 0.40 & 0.51 & 0.48 & 0.43 & 0.38 & 0.38 & 1 \\
0.40 & 0.60 & 0.46 & 0.43 & 0.39 & 0.34 & 0.34 & 0.28 & 1 \\
0.60 & 0.80 & 0.34 & 0.31 & 0.29 & 0.25 & 0.25 & 0.20 & 0.18 & 1 \\
0.80 & 1.00  & 0.21 & 0.20 & 0.18 & 0.16 & 0.15 & 0.13 & 0.11 & 0.08 & 1 \\
1.00 & 2.00   & 0.23 & 0.21 & 0.19 & 0.17 & 0.17 & 0.14 & 0.12 & 0.09 & 0.06 & 1 \\
\end{tabular}
}
\end{center}
\label{tab:CorrelPhistar}
\end{table}

\clearpage
\bibliographystyle{LHCb}
\bibliography{ZeeBib}

\end{document}